\documentclass[journal]{IEEEtran}

\usepackage[colorlinks]{hyperref}

\usepackage{booktabs}
\usepackage{tabularx}

\usepackage{amssymb}
\usepackage{amsbsy}
\usepackage{amsmath}
\usepackage{bm}
\usepackage{verbatim}
\usepackage{mathrsfs}
\usepackage{amsfonts}
\usepackage{graphicx}
\usepackage[tight,footnotesize]{subfigure}
\usepackage[10pt]{moresize}
\usepackage{array}
\usepackage{color}
\usepackage{epsfig}
\usepackage{stfloats}
\usepackage{balance}

\usepackage{hyperref} 

\usepackage{amssymb}

\usepackage[noend]{algpseudocode}
\usepackage{algorithmicx,algorithm}

\usepackage{setspace}
\usepackage{cases}

\usepackage{graphicx}
\usepackage{epstopdf}
\usepackage{multirow}
\usepackage{extarrows}

\usepackage{pifont} 
\usepackage{enumerate}
\usepackage{enumitem}

\newcommand{\subparagraph}{}
\usepackage{titlesec}
\titlespacing{\section}{0pt}{2 ex plus .0ex minus .0ex}{1ex plus .0ex}
\titlespacing{\subsection}{0pt}{1.5 ex plus .0ex minus .0ex}{0.8 ex plus 0.0ex}
\titlespacing{\subsubsection}{0pt}{0.5ex plus .0ex minus .0ex}{0.0ex plus .0ex}

\usepackage{amsmath} 
\allowdisplaybreaks[4]

\ifCLASSINFOpdf

\else

\fi

\begin{document}

\title{WiNPA: Wireless Neural Processing Architecture}

\author{Sai Xu and Yanan~Du
\vspace{-3mm}
\thanks{S. Xu is with the Department of Electronic and Electrical Engineering, University College London, WC1E 7JE, London, UK.  (e-mail: $\rm sai.xu@ieee.org$). Y. Du is with the School of Electrical and Electronic Engineering, the University of Sheffield, Sheffield S10 2TN, UK. (e-mail: $\rm yanan.du@sheffield.ac.uk$)}
}

\maketitle
\begin{abstract}
This article presents a wireless neural processing architecture (\textit{WiNPA}), providing a novel perspective for accelerating edge inference of deep neural network (DNN) workloads via joint optimization of wireless and computing resources. WiNPA enables fine-grained integration of wireless communication and edge computing, bridging the research gap between wireless and edge intelligence and significantly improving DNN inference performance. To fully realize its potential, we explore a set of fundamental research issues, including mathematical modeling, optimization, and unified hardware--software platforms. Additionally, key research directions are discussed to guide future development and practical implementation. A case study demonstrates WiNPA’s workflow and effectiveness in accelerating DNN inference through simulations.
\end{abstract}

\IEEEpeerreviewmaketitle

\section{Introduction}\label{sec:introduction}
\IEEEPARstart{T}HE rise of artificial intelligence (AI) is propelling its widespread integration across various industries, with communication being no exception. At the intersection of communication and AI, research has followed two opposing directions. The first, “AI-empowered communication”, leverages AI techniques to enhance communication performance and optimize resource management~\cite{Letaief2022Edge}, such as semantic communications~\cite{Huang2023Toward} and learning-based wireless communications~\cite{Chen2024Distributed}. The second, “communication-empowered AI”, exploits communication networks to facilitate AI execution, such as edge computing~\cite{Li2020Demand, Deng2020Edge} and over-the-air computation~\cite{Yang2020Federated}. To date, the latter has largely relied on high-level abstractions for system modeling, often overlooking lower-level factors such as wireless channel fading, deep neural network (DNN) operators, or chip microarchitecture~\cite{Li2023Throughput, Zeng2021CoEdge}, which can result in inefficient utilization of wireless and computing resources. Motivated by this, we propose \textit{WiNPA}, a wireless neural processing architecture that represents a lower-level perspective—specifically integrating wireless-level communication and hardware-level computing—to accelerate edge inference of DNN workloads.

As a finer-grained, integrated communication–computation paradigm, WiNPA fundamentally rethinks how wireless networks and edge intelligence platforms can be more tightly orchestrated to enable more efficient DNN inference. Rather than treating wireless transmission and inference as separate steps—where data are first transmitted and then processed~\cite{Shi2020Communication}—WiNPA integrates them into a unified system. As a dedicated wireless infrastructure, this architecture is particularly valuable in many real-world edge inference scenarios. For example, in distributed acoustic networks, the promising schemes of “distributed sensing + edge computing” often increase latency from the ideal tens of milliseconds to several hundred milliseconds. In contrast, WiNPA, through lower-level integration of communication and computation, can utilize resources more efficiently, thereby accelerating edge inference.

Although WiNPA holds great promise, the technical integration of wireless-level communication and hardware-level computing remains largely unexplored, presenting several fundamental challenges. First, developing realistic mathematical models to explore system performance limits is difficult, as inference performance depends on computation and communication resources, transceiver specifications, and wireless conditions. Existing models consider only part of these factors, leading to discrepancies with real-world performance. Second, formulating and solving optimization problems under practical constraints is highly complex. Edge DNN inference entails tightly coupled computation and communication, dynamic wireless environments, and multi-node coordination, all of which make efficient algorithm design particularly challenging. Third, achieving hardware–software co-design for wireless-accelerated edge inference is nontrivial. Current platforms are usually designed independently, and building scalable, unified systems requires system-level optimization to realize the expected performance gains.

These fundamental challenges motivate us to bridge the gap between conceptual ideas and practical implementation. To this end, we discuss critical issues such as mathematical modeling, joint optimization of communication and computation resources, and hardware–software co-design, offering researchers a fine-grained perspective on actionable and forward-looking research directions. Furthermore, we identify several critical technical areas that warrant further investigation, including hardware microarchitecture, emulator development, scheduling algorithms, inference-oriented wireless communication, and end-to-end toolchain. Finally, we present a case study of wireless-accelerated edge inference to illustrate the construction of a mathematical model and the execution of WiNPA for DNN inference workloads.

\section{Research Framework}\label{sec:framework}
This section presents concrete research pathways to tackle the aforementioned fundamental challenges, providing practical insights and guidance for future investigations.

\subsection{Mathematical Modeling}\label{ssec:model}
\begin{figure*}
\centering
\includegraphics[width= 7.2 in]{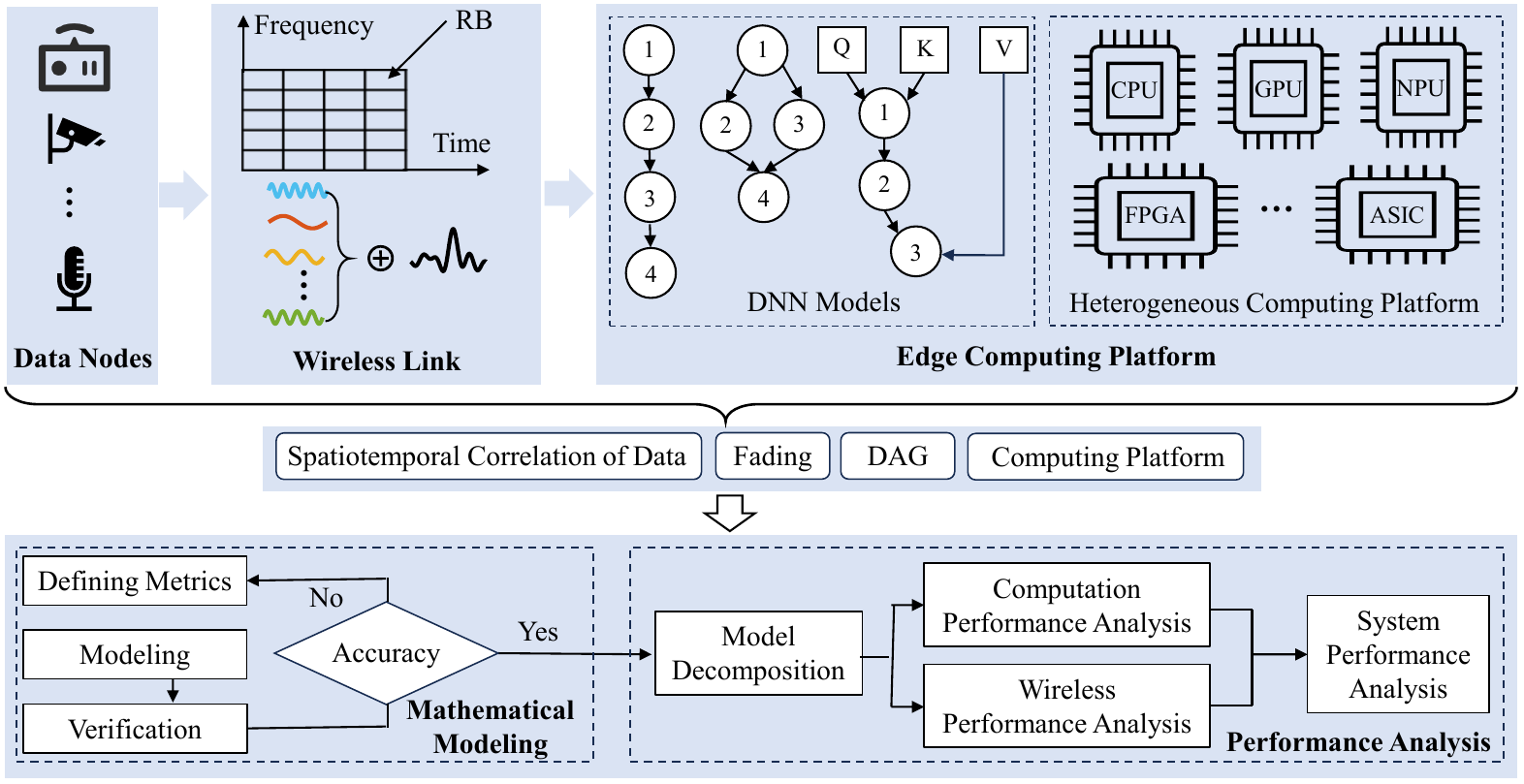}
\caption{An envisioned framework for mathematical modeling and performance analysis.}
\label{Fig1}
\end{figure*}
Mathematical modeling underpins the design and optimization of wireless-accelerated edge inference systems. WiNPA moves beyond traditional isolated designs to capture the critical interactions of wireless communication, parallel computing, and the data dependencies of DNN inference tasks. The process begins with defining a representative system model that reflects how the input data is acquired, transmitted, and processed at the edge. A comprehensive set of performance metrics—including throughput, latency, power \& energy efficiency, and inference accuracy—is then established to quantify system behavior. By integrating principles from wireless communication and parallel computing, a unified model describes how hardware characteristics, wireless link dynamics, and the spatiotemporal correlations in the input data collectively influence system performance. Additionally, simulation and experimental validation should be conducted to ensure that the model not only explains observed behavior but also guides the design of more efficient wireless-accelerated edge inference systems.

Building upon the mathematical model, further performance analysis provides valuable insights into the system’s operational limits and guides potential optimization strategies. The system is decomposed into two key modules—communication and computation—to assess each component’s contribution to the overall performance. Computational capacity, memory bandwidth, communication resource, and wireless channel characteristics are jointly considered to quantify achievable performance under varying DNN workloads and input data types. Moreover, the spatiotemporal correlations of input data are modeled statistically to estimate communication efficiency, while the joint analysis of computation and communication reveals potential bottlenecks in co-designed systems. Fig.~\ref{Fig1} illustrates an envisioned framework for mathematical modeling and performance analysis.

\subsection{Optimization}\label{ssec:optimization}
For unimodal tasks, the optimization process of WiNPA starts by defining clear performance objectives and fundamental constraints on wireless and computational resources, while incorporating the spatiotemporal correlations of input data. A corresponding task-scheduling model is then established, and an appropriate scheduling strategy—either static or dynamic—is selected according to system characteristics. Efficient optimization algorithms are indispensable to enhance overall system performance \cite{Zhou2025TaiChi}. Finally, the optimization framework is refined by improving objective or cost function design, thereby increasing its practicality for complex real-world applications.

\begin{figure*}
\centering
\includegraphics[width= 7.2 in]{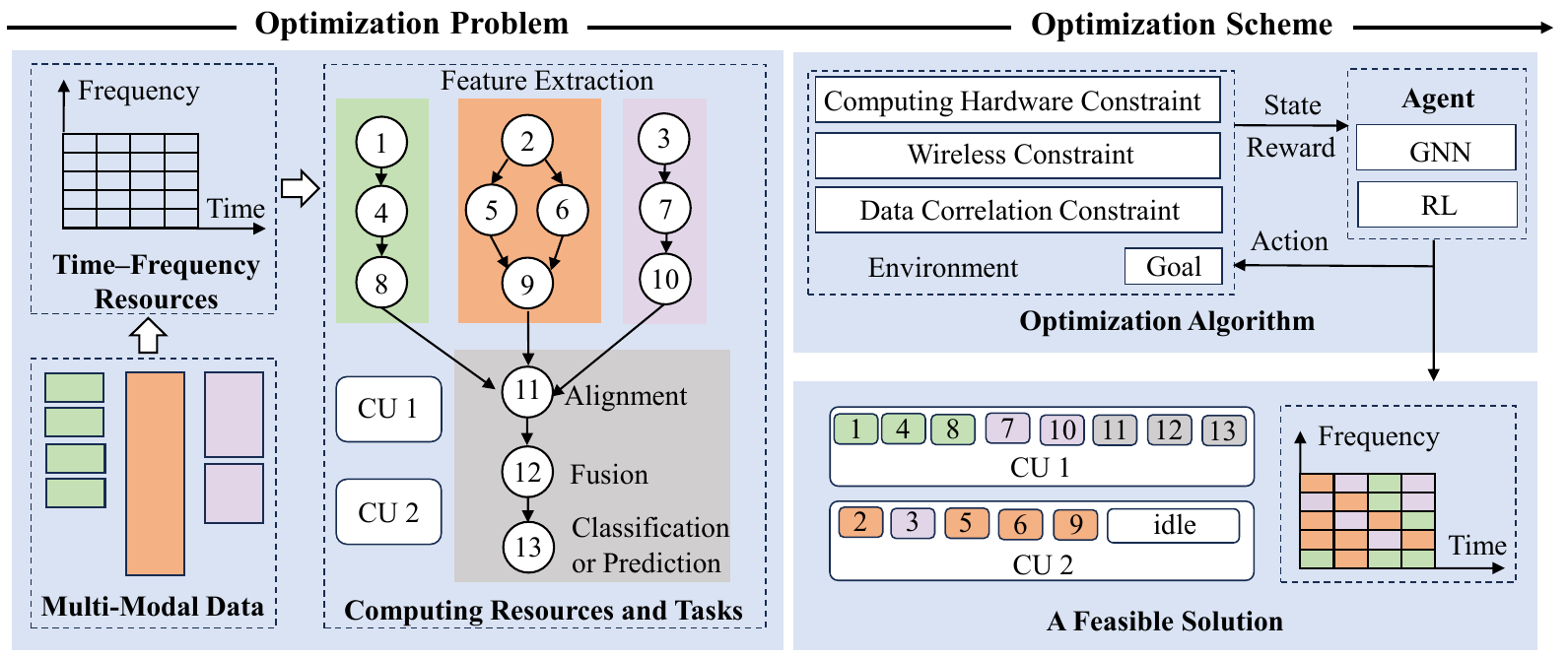}
\caption{A GNN-RL-based optimization method for mapping DNN inference tasks to CUs.}
\label{Fig2}
\end{figure*}

For multimodal tasks, WiNPA extends the unimodal approach by incorporating additional communication and computation characteristics specific to multimodal systems, including hardware constraints, communication mechanisms, and resource allocation policies. Data dependencies across modalities are represented using spatiotemporal correlation graphs, while advanced optimization techniques are employed to maximize system-level computational efficiency through intelligent scheduling. The proposed multimodal optimization strategies are rigorously evaluated and iteratively refined, ensuring robust performance across complex, real-world edge intelligence deployments. Fig. \ref{Fig2} envisions a graph neural network (GNN) and reinforcement learning (RL)-based optimization method that maps DNN inference tasks onto computing units (CUs) for acceleration.
\subsection{Unified Hardware–Software Platforms}\label{ssec:platfor}
Efficient wireless-accelerated edge inference systems cannot rely on fragmented designs, where computation and communication modules are developed independently and integrated only at a late stage. Such fragmented platforms often suffer from sequential delays, poor coordination, and suboptimal performance under dynamic channels and complex workloads. WiNPA represents a fully unified hardware-software platform that addresses these limitations by enabling joint design across hardware microarchitecture, compilers, task scheduling, and communication protocols. This integrated approach accounts for wireless channel variability, resource constraints, and data dependencies from the outset, while optimizing pipeline parallelism of the communication and computation modules, caching, memory access, and data flows to enhance throughput and responsiveness. Unified validation ensures seamless coordination between computation and communication, guaranteeing robust performance in complex edge scenarios. 

\begin{figure*}
\centering
\includegraphics[width= 7.2 in]{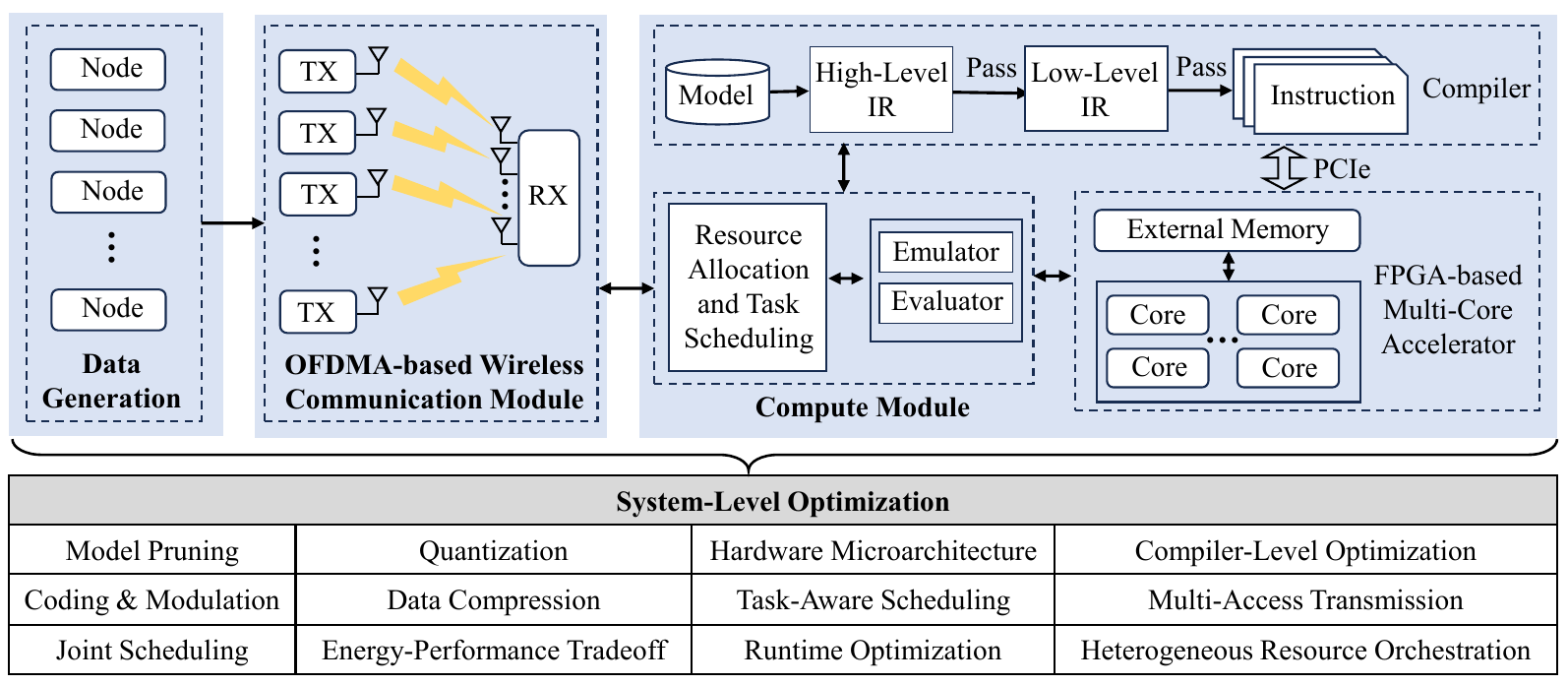}
\caption{An architecture for the unified hardware–software platform and typical system-level optimization approaches.}
\label{Fig3}
\end{figure*}

Beyond hardware integration, the platform implements system-level optimization strategies. Resource interfaces, data paths, and scheduling mechanisms are explicitly modeled to maximize efficiency across compute and communication modules. Task graphs capture DNN inference characteristics—including computation intensity and communication dependencies—revealing opportunities for spatial and temporal parallelism. Joint optimization strategies balance computational load, bandwidth, and energy consumption, achieving global efficiency in scheduling and resource allocation. End-to-end simulations and real-world testing under varying channel conditions guide iterative improvements, while predictive analytics and adaptive runtime management allow dynamic reconfiguration of resources. Together, these strategies enable low-latency, energy-efficient, and robust edge DNN inference, laying the foundation for scalable, intelligent deployment across heterogeneous multi-node edge systems. Fig. \ref{Fig3} envisions an architecture for the unified hardware–software platform and typical system-level optimization approaches.
\section{Important Research Directions}\label{sec:agendas}

\subsection{Hardware Microarchitecture}\label{sec:hardware}
Tight integration between wireless communication and edge computing requires a fundamental redesign of hardware microarchitecture to support coordinated processing, data flow optimization, and seamless resource sharing. One key approach is the co-design of wireless signal processing and computational hardware. By reusing computing and memory resources between communication tasks—such as filtering and demodulation—and AI acceleration, the total processing delay can be reduced and on-chip storage utilized more efficiently. Another strategy is dynamic, data-driven computation, where large input datasets are partitioned and processed as soon as partial data arrives, mitigating the effects of wireless channel variability and enabling computation and communication to proceed in parallel. Additionally, workload-aware resource allocation can further boost performance: computation-intensive tasks are assigned more CUs, while memory-intensive tasks benefit from optimized caching and data prefetching.

Beyond these case-specific examples, more generalizable design principles are equally important. Modular, reconfigurable accelerators can adapt to varying communication and computation patterns, hierarchical scheduling frameworks can coordinate multiple heterogeneous nodes, and predictive hardware-software co-design can leverage runtime network and workload statistics for dynamic resource management. Together, these strategies point toward a flexible, communication-aware, and computation-adaptive microarchitecture, capable of delivering low-latency, energy-efficient, and robust edge inference performance across diverse real-world deployments.

\subsection{Emulator}\label{sec:scheduling}
WiNPA integrates wireless transmission with edge computing, where end-to-end performance—latency, energy, and reliability—is shaped by data generation, transmission, and task computation. Evaluating the impact of network protocols, data placements, accelerator architectures, or scheduling policies requires simulation or emulation. However, the current ecosystem is fragmented: wireless simulators (e.g., ns-3, OMNeT++) model channels, bandwidth, and energy but provide limited support for DNN operator behaviors or data reuse; accelerator simulators (e.g., Timeloop, MAGMA~\cite{Kao2022MAGMA}, ScaleSim) focus on computation and memory, often assuming immediate data availability. The absence of unified time semantics, standard data interfaces, and coupled energy models limits accurate assessment of the interactions across data, communication, and computation.

To address this gap, future emulators should evolve into deeply integrated, multi-scale platforms that jointly model communication and computation. Architecturally, an event-driven core can coordinate data, wireless transmission, queue scheduling, and computation tasks. At the module level, high-fidelity components can be embedded: wireless links via ns-3 or OMNeT++, and accelerator computations via Timeloop or MAESTRO. Key research directions include: (i) multi-timescale synchronization—aligning millisecond-scale wireless events with cycle-level computations; (ii) end-to-end energy and bandwidth co-modeling—enabling closed-loop analysis of per-bit energy versus computation latency for system-level optimization; and (iii) co-design space exploration (Co-DSE)—simultaneously optimizing wireless parameters (bandwidth, scheduling, batching) and accelerator parameters \cite{Zeng2022Serving} (array size, buffer hierarchy, mapping strategies). Such an integrated simulator enables systematic evaluation of design choices on latency, energy efficiency, and reliability, providing quantitative guidance for cross-module optimization and supporting the development of edge-intelligent computing systems.

\subsection{Scheduling Algorithms}\label{sec:scheduling}
In WiNPA, nodes primarily sense environmental data and transmit it to a central computing node over wireless links for DNN inference. Joint scheduling in this scenario aims to optimize both wireless communication and central computing resources simultaneously. The system monitors each node’s data generation rate, wireless bandwidth, latency, and channel quality, as well as the load, available computing capacity, and task queue lengths at the computing center. Based on this information, scheduling determines each node’s upload rate, transmission window, and priority to prevent central overload and wireless congestion. For example, time slots can be dynamically allocated according to data volume and link conditions, or data can be compressed or transmitted in batches under poor network conditions to reduce latency and energy consumption. End-to-end simulations incorporating wireless models, data generation patterns, and inference delays can evaluate scheduling performance in terms of delay, throughput, and resource utilization.

Efficient joint scheduling requires coordinated optimization of both transmission and computation. RL can learn optimal strategies in dynamic network and central load scenarios, determining node upload order, rate, and priority, while adapting task allocation to central queue states. GNNs can model network topology, link status, and central load to predict bottlenecks, providing precise state inputs for RL. Heuristic algorithms quickly generate feasible solutions under short-term fluctuations, balancing computation and communication resources. Additional optimizations include data compression or feature extraction, batch uploads to reduce queue pressure, asynchronous inference to utilize compute units, and prioritizing latency-sensitive tasks. These strategies enable concurrent multi-node uploads with low latency, high throughput, and efficient resource utilization, providing a robust solution for integrated edge sensing and centralized inference.

\subsection{Inference-Oriented Wireless Communication}\label{sec:wireless}
In edge computing systems, traditional wireless communication methods often fail to meet the latency and dependency requirements of task-specific data. Data for DNN inference typically exhibits temporal or spatial dependencies, meaning certain information cannot be processed until preceding data arrives. Simple first-come-first-served or uniform scheduling can create bottlenecks at the computing center and increase overall latency. To address this, communication strategies must account for data importance, dependency order, and node generation rates. Critical-path data can be allocated priority transmission channels, with dynamic adjustments to upload rates and time windows, while non-critical or loosely dependent data may be delayed or batched to reduce wireless congestion. Task dependency graphs or hierarchical dependency models can be integrated with communication resource allocation to achieve end-to-end scheduling optimization.

Optimization strategies can further enhance wireless transmission efficiency for DNN inference. Edge nodes may perform feature extraction, data compression, sparse encoding, or selective uploads to transmit only the most relevant information, reducing network load. Channel coding techniques, such as adaptive forward error correction or rateless codes, can improve transmission reliability while minimizing retransmissions. In multi-node scenarios, bandwidth can be dynamically allocated based on task dependency graphs and node load, and priority-based scheduling or multi-access schemes (e.g., OMA/NOMA) ensure critical data reaches the central node first. Predictive models, such as GNN, can forecast node data generation rates and link conditions, enabling proactive adjustment of transmission order, coding rates, and channel assignment. By jointly considering data dependencies, task topology, channel coding, and network state, specialized wireless communication schemes for AI inference can achieve low latency, high throughput, and efficient spectrum utilization, significantly improving the performance and responsiveness of edge computing systems.

\subsection{End-to-End Toolchain}\label{sec:toolchain}
WiNPA demands a seamless integration of wireless transmission and edge computing. A communication-computation co-design toolchain spans the full workflow—from data acquisition and edge preprocessing, through wireless transmission, to central inference and result delivery. Core components include an end-to-end compiler that optimizes execution across heterogeneous hardware, model compression and quantization for pruning, low-bit representation, or hierarchical partitioning, and a task scheduler that dynamically allocates upload sequences and compute resources based on task dependencies, node load, and data rates. On the communication side, channel coding, adaptive modulation, batch transmission, and priority scheduling ensure critical data reaches the central node reliably and with minimal latency, while predictive modules anticipate network conditions and node workloads to proactively adjust scheduling. Together, these mechanisms enable synchronized optimization of both transmission and computation in real-world edge computing scenarios.

Beyond the core modules, a practical co-design toolchain emphasizes scalability for multi-node, multi-task deployments, reconfigurability to adapt model partitioning and scheduling to dynamic network and compute conditions, automation and intelligence through compiler, compression, and predictive optimizations, and observability to monitor links, workloads, and system health in real time. Fault tolerance and robustness further ensure continuous operation under link fluctuations or node failures. By unifying communication and computation in a single, adaptive framework, the toolchain delivers low latency, high throughput, and efficient resource utilization—making it a deployable, end-to-end solution for next-generation edge computing systems.
\begin{figure}
\centering
\includegraphics[width= 3.5 in]{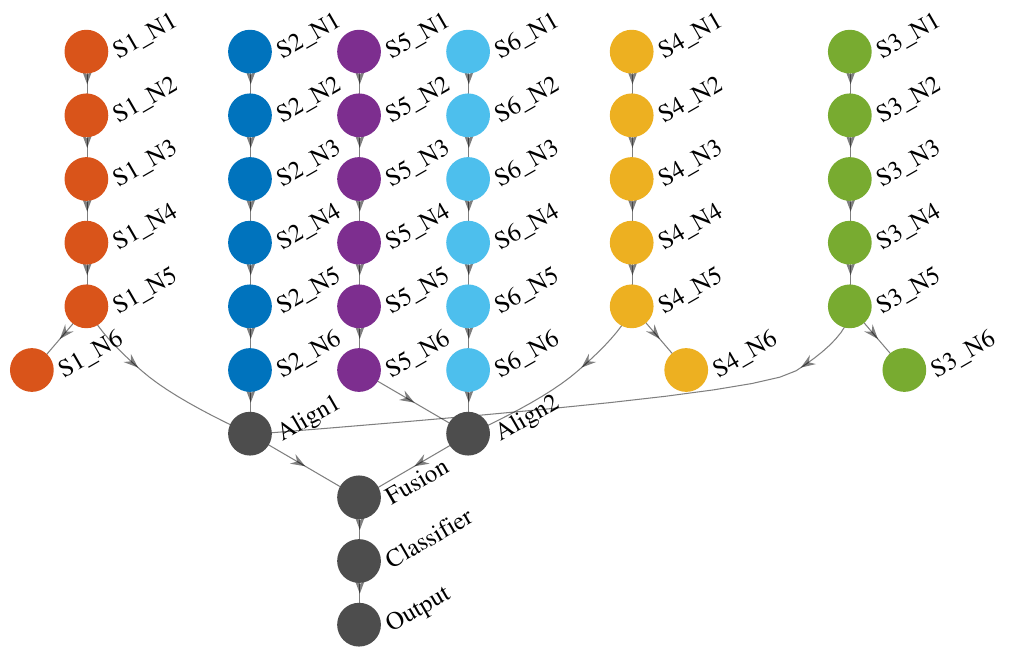}
\caption{A random DAG for the simulation.}
\label{Fig4}
\end{figure}
\begin{figure*}[t] 
\centering
\includegraphics[width= 7.2 in]{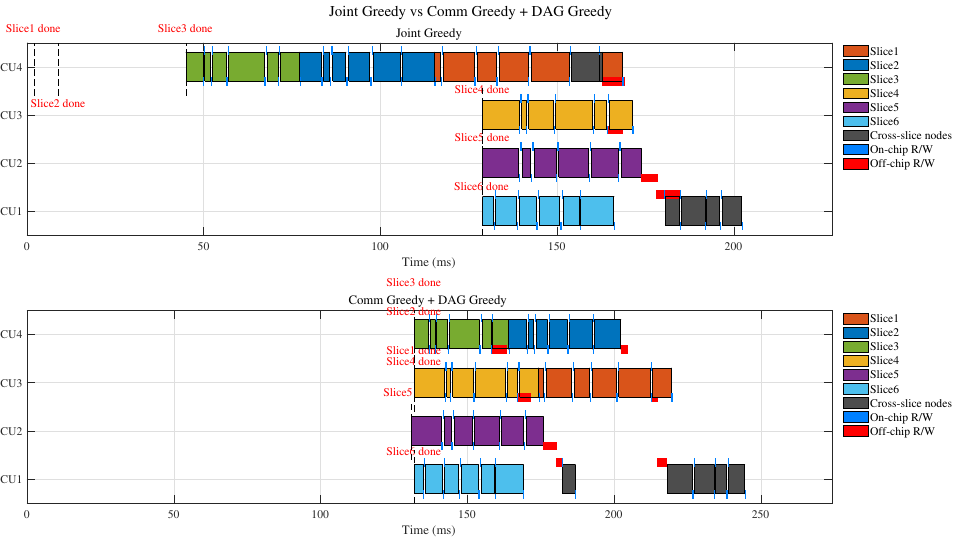}
\caption{Simulation results.}
\label{Fig5}
\end{figure*}
\section{Case Study}\label{sec:case}
This section presents a simple case study to demonstrate the execution of WiNPA for wireless-accelerated edge inference, with simulation results validating the performance gain achieved through the communication-computation co-design.

\subsection{System Overview}\label{sec:overview}
Consider a wireless-accelerated edge inference system that consists of three core components: communication, computation, and a computational graph.
In this system, $K$ nodes each generate a slice of modal data. Each slice must be completely received by the edge computing center before processing can begin, while the slices are computationally independent of one another. The data are transmitted to the edge computing center via orthogonal frequency-division multiple access (OFDMA) over wireless channels. The available wireless resources are divided into $T \times F$ time-frequency resource blocks (RBs), which are allocated among the $K$ nodes. The edge computing center is equipped with $M$ heterogeneous CUs to process the incoming data. Upon arrival at the edge computing center, each modality is processed through its corresponding branch of the computational graph for feature extraction. The resulting features are then aligned across modalities and fused using strategies such as concatenation or attention-based mechanisms. The fused representation is finally used for tasks such as classification or prediction. The execution process can be represented by a directed acyclic graph (DAG), where nodes denote computation tasks and edges represent data dependencies. The overall workflow includes slice-level data transmission, intra-slice computation, cross-slice fusion, and final classification  or prediction. This design enables parallel processing of modalities and leverages the heterogeneous CUs of the edge center to accelerate multi-modal inference.

\subsection{Simulation Setup}\label{sec:setup}
The simulation parameters are set as follows. The numbers of slices and CUs are set to $N_\text{s} = 6$ and $N_\text{CU} = 4$.
Each slice \( s \) has an associated data payload \( D_s \) that must be fully transmitted before computation starts with $\mathcal{D} = [1,\, 7.5,\, 25,\, 10,\, 35,\, 50]~\text{kB}$. The wireless channel fluctuates over time to emulate dynamic fading conditions. The instantaneous throughput of each RB at time slot \( t \) and frequency index \( f \) for different slices follows $R_{t,f} \sim \mathcal{U}(R_{\min}, R_\text{max})$, where \( R_\text{min} = 20 \, \text{kbps} \) and \( R_\text{max} = 2000 \, \text{kbps} \). The system operates over \( F = 8 \) frequency resources and a simulation horizon of \( T = 500 \) time slots, with each time slot measured in milliseconds.  During simulation, each RB is dynamically allocated to one slice per time slot, transferring \( R_{t,f}/(8 \times 1000) \) kb of data. For each slice $s$, we define a DAG $\mathcal{G}_s = (\mathcal{V}_s, \mathcal{E}_s)$, where $\mathcal{V}_s$ is the set of computation nodes and $\mathcal{E}_s$ is the set of directed dependencies. The number of nodes per slice is randomly chosen between 5 and 6 to emulate heterogeneous task structures. Additionally, several cross-slice dependencies are introduced to model data fusion and alignment stages (e.g., “Align1”, “Align2”, “Fusion”, and “Classifier”). These cross-links ensure that the system mimics multi-modal AI pipelines such as collaborative perception or multi-sensor fusion networks. Each operator node \( v \in \mathcal{V} \) is characterized by a set of timing parameters:
\[
t_{\mathrm{comp}}(v), \; 
t_{\mathrm{read\_on}}(v), \; 
t_{\mathrm{write\_on}}(v), \;
t_{\mathrm{read\_off}}(v), \;
t_{\mathrm{write\_off}}(v),
\]
representing computation time, on-chip read/write delays, and off-chip read/write delays, respectively.
To emulate runtime variability, these parameters are sampled from the following uniform ranges: $t_{\mathrm{comp}} \in [5, 25]~\text{ms}$, 
$t_{\mathrm{read\_on}}, t_{\mathrm{write\_on}} \in [0.1, 0.6]~\text{ms}$, $t_{\mathrm{read\_off}}, t_{\mathrm{write\_off}} \in [2, 8]~\text{ms}$.
The total end-to-end latency is adopted as performance metric, which is given by
\[
T_{\mathrm{total}} = \max_{v \in V}~t_{\mathrm{finish}}(v),
\]
where \( t_{\mathrm{finish}}(v) \) is the completion time of node \( v \).
\subsection{Simulation Results}\label{sec:result}
Fig.~\ref{Fig4} shows a randomly generated DAG and Fig.~\ref{Fig5} depicts the simulation results, comparing two scheduling strategies. For simplicity, each node operates in a strict read–compute–write sequence. \texttt{Joint Greedy} refers to the joint optimization of both communication and computation resource allocation as well as task scheduling. 
At each time slot, the algorithm evaluates the estimated end-to-end latency impact of assigning a RB to each slice and greedily selects the slice that minimizes the overall system completion time. 
\texttt{Communication Greedy + DAG Greedy} refers to decoupled communication and computation scheduling. 
The communication stage greedily allocates RBs to the slice with the largest remaining data size, 
while DAG tasks are scheduled independently using a topological greedy approach once their corresponding data transmissions complete.
According to the simulations, the latency of \texttt{Joint Greedy} and \texttt{Communication Greedy + DAG Greedy} are 202.23 ms and 244.48 ms, respectively. 
Both strategies are compared in terms of \(T_{\mathrm{total}}\), with Gantt charts visualizing CU utilization and task overlap, 
which reveals that joint optimization significantly accelerates inference.

\section{Conclusions}\label{sec:conclusions}
This article presented WiNPA, providing a novel perspective for accelerating edge inference of DNN workloads through the tight integration of
wireless-level communication and hardware-level computing. A case study illustrated the implementation of mathematical modeling and deep co-optimization, with preliminary simulations validating its effectiveness in improving inference performance. These results highlighted both the potential and necessity of low-level communication–computation co-optimization, and provided a foundation for the practical deployment of wireless-accelerated edge inference systems.

\ifCLASSOPTIONcaptionsoff
  \newpage
\fi

\end{document}